\documentclass{birkjour}
\usepackage [latin1]{inputenc}

\theoremstyle{definition}

\theoremstyle{remark}

\numberwithin{equation}{section}

\newcommand{\BibTeX}{B\kern-0.1emi\kern-0.017emb\kern-0.15em\TeX}
\newcommand{\XYpic}{$\mathrm{X\kern-0.3em\raisebox{-0.18em}{Y}}$-$\mathrm{pic}\,$}

\newcommand{\cl}{C \kern -0.1em \ell}  

\newcommand{\ed}{\end{document}}
\begin{document}
\title[Pseudosymmetric spacetime to be a prefect fluid spacetime]
{Sufficient conditions for a pseudosymmetric spacetime to be a perfect fluid spacetime}

\author[P. Zhao , U. C. De, B. \"{U}nal  and  K. De]
{Peibiao Zhao , Uday Chand De, B\"ulent \"Unal and  Krishnendu De}

\address
{ Department of Applied Mathematics, Nanjing University of Science and Technology, Nanjing 210094, P. R. China.}
\email {pbzhao@njust.edu.cn}

\address
{Department of Pure Mathematics, University of Calcutta
35, Ballygaunge Circular Road
Kolkata 700019, West Bengal, India.}
\email {uc$_{-}$de@yahoo.com}
\address{Department of Mathematics, Bilkent University,
Bilkent, 06800 Ankara, Turkey}
\email{bulentunal@mail.com}
\address
 { Department of Mathematics,
 Kabi Sukanta Mahavidyalaya,
The University of Burdwan.
Bhadreswar, P.O.-Angus, Hooghly,
Pin 712221, West Bengal, India.}
\email{krishnendu.de@outlook.in }

\footnotetext {$\bf{2010\ Mathematics\ Subject\ Classification\:}.$ 53D10, 53D15, 53D25, 53D35.
\\ {Key words and phrases: Lorentzian manifolds, perfect fluid spacetime, $GRW$ spacetimes, pseudosymmetric spacetimes\\
}}

\begin{abstract}
The aim of the present paper is to obtain the condition under which a pseudosymmetric spacetime to be a perfect fluid spacetime. It is proven that a pseudosymmetric generalized Robertson-Walker spacetime is a perfect fluid spacetime. Moreover, we establish that a conformally flat pseudosymmetric spacetime is a generalized Robertson-Walker spacetime. Next, it is shown that a pseudosymmetric dust fluid with constant scalar curvature satisfying Einstein's field equations without cosmological constant is vacuum. Finally, we construct a non-trivial example of pseudosymmetric spacetime.
\end{abstract}
\maketitle
\numberwithin{equation}{section}
\newtheorem{theorem}{Theorem}[section]
\newtheorem{lemma}[theorem]{Lemma}
\newtheorem{proposition}[theorem]{Proposition}
\newtheorem{corollary}[theorem]{Corollary}
\newtheorem*{remark}{Remark}
\newtheorem{Agreement}[theorem]{Agreement}
\newtheorem{definition}[theorem]{Definition}
\newtheorem{example}[theorem]{Example}

\section{Introduction}

Let $M$ be a semi-Riemannian manifold of dimension $n\ge2$ equipped with a semi-Riemannian metric $g$ of signature $(m, p)$, where $m+p=n$. If $g$ is a Lorentzian metric of signature $(1, n-1)$ or $(n-1, 1)$, then $M$ endowed with $g$ is said to be an $n$-dimensional Lorentzian manifold \cite{neil}. If $M=-I \times_{ \mathfrak{f}}\,\mathcal{M}$, where $I$ is an open interval of $\mathbb{R}$ (set of real numbers), $\mathcal{M}$ is a Riemannian manifold of dimension $n-1$ and $\mathfrak{f}>0$ denotes a smooth function, named as warping function or scale factor, then $M$ is said to be a generalized Robertson-Walker $(\text{briefly}, GRW)$ spacetime (\cite{alias1}, \cite{alias2}).  Particularly, if we suppose that $\mathcal{M}$ is a Riemannian manifold of dimension $3$ with constant scalar curvature, then the $GRW$ spacetime becomes a Robertson-Walker ($RW$) spacetime. This infers that $GRW$ spacetimes are the natural extension of $RW$ spacetimes. It is well-known that the Lorentzian Minkowski spacetime, the Friedmann cosmological models, the static Einstein spacetime, the Einstein-de Sitter spacetime, the de Sitter spacetime are included in the $GRW$ spacetimes \cite{sanches1}. They have also many applications in inhomogeneous spacetimes admitting an isotropic radiation. Several remarkable results on  $GRW$ spacetimes are investigated in (\cite{des2}, \cite{chaubey2}, \cite{bychen},  \cite{manticasuhde}, \cite{survey}, \cite{manticamolinaride}) and by others.

A vector field $\zeta$ on an $n$-dimensional semi-Riemannian manifold $M$ is said to be concircular if it satisfies $\nabla_{X} \zeta=\aleph\, X$ for some smooth function $\aleph$, where $\nabla_{X}$ denotes the covariant derivative along the smooth vector field $X$ of $M$. The notion of concircular vector field on Riemannian manifold has been introduced by Failkow \cite{fail}, and further studied by \cite{chen2016} and others.

 Chen \cite{bychen} has given a local characterization of a Lorentzian manifold admitting a timelike concircular vector field:{\vspace{0.3cm}}\\
{\bf Theorem A.} \cite{bychen} A Lorentzian manifold $M$ of dimension $n \ge 3$ is a $GRW$ spacetime if and only if it admits a timelike concircular vector field such that $X_{i,j}=\rho g_{ij}$ for some smooth function $\rho$. {\vspace{0.3cm}}\\
The concept of torse-forming vector field is weaker than the concircular one, but with an additional condition, the following local characterization of $GRW$ spacetimes is possible.{\vspace{0.3cm}}\\
{\bf Theorem B.} \cite{survey} A Lorentzian manifold of dimension $n \ge 3$ is a $GRW$ spacetime if and only if it admits a unit timelike vector, $u_{i,j}=\varphi(g_{ij}+u_{i}u_{j})$, that is also an eigenvector of the Ricci tensor.{\vspace{0.3cm}}\\
An $n$-dimensional spacetime $M$ with a non-vanishing Ricci tensor $R_{ij}$ is said to be a perfect fluid spacetime if $R_{ij}$ satisfies the relation
\begin{equation}
\label{pfs}
R_{ij}=\alpha g_{ij}+\beta A_{i} A_{j}
\end{equation}
for some smooth functions $\alpha$ and $\beta$ on $M$. Here $g$ denotes the Lorentzian metric of $M$ and $A_{i}$ is the velocity vector such that $g_{ij}A^{i}A^{j}=-1$ and $A_{i}=g_{ij}A^{j}$. It is noticed that a $RW$ spacetime is a perfect fluid spacetime (\cite{neil}, Theorem 12.11). Every $GRW$ spacetime with $n=4$ is a perfect fluid spacetime if and only if it is a $RW$ spacetime. 
 We cite (\cite{blaga2}, \cite{blaga1}, \cite{de2021}, \cite{malick},  \cite{ucde2}, \cite{survey}, \cite{mantica2019}, \cite{foz1}) and its references for some deep results of the perfect fluid spacetimes.

During the investigation of conformally flat Riemannian manifolds of class one, Sen and Chaki \cite{sen1} found that the covariant derivative of the curvature tensor $R$ of type $(0, 4)$ satisfies
\begin{equation}
\label{1}
R_{hijk, l}=2v_{l} R_{hijk}+v_{h}R_{lijk}+v_{i}R_{hljk}+v_{j}R_{hilk}+v_{k}R_{hijl},
\end{equation}
where ",  (comma)" denotes the covariant derivative with respect to the metric tensor, $R_{hijk}$ are the components of the curvature tensor and $v_{l}$ is a non-zero covector. Later, Chaki \cite{chaki2} named a Riemannian manifold to be pseudosymmetric if the curvature tensor satisfies the condition (\ref{1}). Then Chaki and De \cite{chaki1} examined the Riemannian manifolds with the above condition. If the covector $v_{l}$ vanishes, then the manifold becomes a locally symmetric manifold in the sense of Cartan. An $n$-dimensional pseudosymmetric manifold is denoted by $(PS)_n$. In $1992$, Deszcz \cite{des1} introduced the notion of a pseudosymmetric manifold which is different from Chaki's notion. According to Deszcz, a pseudo-symmetric manifold is said to be pseudosymmetric if $Riem \cdot Riem$ and $Q(g, Riem)$ are linearly dependent at each point of the manifold, where $Riem$ and $Q$ denote the Riemannian curvature and Ricci operator, respectively. Thus we have
$Riem \cdot Riem=\mathfrak{h}Q(g, Riem)$ for some smooth function $\mathfrak{h}$. This paper deals with Chaki's notion of pseudosymmetric. Pseudosymmetric manifolds have been studied by several authors such as (\cite{de1}, \cite{degazi}, \cite{mansuh}, \cite{fo1}) and others.

In \cite{prv1}, Prvanovi\'{c} called such a manifold as a generalized recurrent manifold and proved that the associated covector $v_{l}$ is gradient, that is, irrotational.

On the other hand, a spacetime of general relativity is regarded as a $4$-dimensional time-oriented Lorentzian manifold $(M, g)$. 
In general relativity, the matter content of the spacetime is described by the energy momentum tensor. The matter content is assumed to be fluid having density, pressure and possessing dynamical and kinematical quantities like velocity, acceleration, vorticity, shear and expansion. The fluid is called perfect because of the absence of heat conduction terms and stress tensor corresponding to viscosity. For a perfect fluid, the energy momentum tensor $T_{ij}$ is given by
\begin{equation}
\label{2}
T_{ij}=(\sigma+p)A_{i}A_{j}+p g_{ij},
\end{equation}
where   $p$ and $\sigma$ are the isotropic pressure and energy density of the perfect fluid, respectively \cite{neil}.

The Einstein's field equations without cosmological constant is given by
\begin{equation}
\label{3a}
R_{ij}-\frac{R}{2}g_{ij}=\kappa T_{ij},
\end{equation}
where $R_{ij}=R^{h}_{ijh}$ is the Ricci tensor, $R=g^{ij}R_{ij}$ denotes the scalar curvature and $\kappa$ is the gravitational constant.
The Einstein's field equations, in fact, connect the distribution of mass (represented by the energy momentum tensor) with the curvature of the spacetime (represented by the Einstein tensor).

In \cite{manticamolinaride}, Mantica, Molinari and De proved that a $GRW$ spacetime with divergence free conformal curvature tensor ($C^{m}_{jkl, m}=0$) is a perfect fluid spacetime.\\
Our aim is to improve the above result of \cite{manticamolinaride}. We  will prove the following:
\begin{theorem}
\label{thm1}
Every pseudosymmetric $GRW$ spacetime is a perfect fluid spacetime.
\end{theorem}
The converse of Theorem \ref{thm1} is not true in general. In this series, Mantica, Molinari and De \cite{manticamolinaride} have established the following theorem:{\vspace{0.2cm}}\\
{\bf Theorem C.} \cite{manticamolinaride} A perfect fluid spacetime together with $div \, C=0$ is a generalized Robertson-Walker spacetime with Einstein fiber, provided that the velocity vector field of the perfect fluid is irrotational.

Motivated from Theorem C, we are going to prove the following theorem:
\begin{theorem}
\label{thm2}
Every conformally flat $(PS)_{n}$ spacetime is a $GRW$ spacetime.
\end{theorem}

\section{Pseudosymmetric $GRW$ spacetimes}
This section is dedicated to give the condition for which a pseudosymmetric spacetime to be a perfect fluid spacetime. {\vspace{0.2cm}}\\
 {\it Proof of Theorem \ref{thm1}.}
Let us suppose that the pseudosymmetric spacetime is a $GRW$ spacetime, then from Theorem {B} we have
\begin{equation}
\label{4}
v_{i,j}=\varphi(g_{ij}+v_{i}v_{j}),
\end{equation}
where $\varphi$ is a scalar. Again the covariant derivative of equation (\ref{4}) gives
\begin{equation}
\label{5}
v_{i,jk}=\varphi_{k}(g_{ij}+v_{i}v_{j})+\varphi\{\varphi(g_{ik}+v_{i}v_{k})v_{j}+\varphi (g_{jk}+v_{j}v_{k})v_{i}\},
\end{equation}
where $\varphi_{k}=\varphi_{, k}$ (the covariant derivative of the smooth function $\varphi$). Interchanging $j$ and $k$ in the above equation and then subtracting the foregoing equation from (\ref{4}) and using Ricci identity, we infer
\begin{eqnarray}
\label{6}
&&v_{h} R^{h}_{ijk}=(\varphi_{k}g_{ij}-\varphi_{j}g_{ik})+v_{i}(\varphi_{k}v_{j}-\varphi_{j}v_{k})+\varphi^2 (g_{ik} v_{j}-g_{ij}v_{k})\nonumber\\&&
\,\,\,\,\,\,\,\,\,\,\,\,\,\,\,\,\,\,\,=(\varphi_{k}-\varphi^2 v_{k})g_{ij}- (\varphi_{j}-\varphi^2 v_{j})g_{ik}+v_{i}\varphi_{k}v_{j}-\varphi_{j}v_{i}v_{k}\nonumber\\&&
\,\,\,\,\,\,\,\,\,\,\,\,\,\,\,\,\,\,\,=w_{k}(g_{ij}+v_{i}v_{j})-w_{j}(g_{ik}+v_{i}v_{k}),
\end{eqnarray}
where $w_{k}=\varphi_{k}-\varphi^2 v_{k}$. Multiplying equation (\ref{6}) with $g^{ij}$, we get
\begin{equation}
\label{7}
v_{h}R^{h}_{k}=(n-2)w_{k}-w_{j} v^{j}v_{k}.
\end{equation}
If $v_{k}$ is an eigenvector of the Ricci tensor, then $w_{k}=f v_{k}$ for some scalar $f$. Hence equation (\ref{7}) turns into
\begin{equation*}
\label{8}
v_{h}R^{h}_{k}=(n-1)f v_{k}.
\end{equation*}
It is noted that the associated vector field $v_{i}$ is gradient \cite{prv1}. Using this fact in equation (\ref{1}), we obtain
\begin{eqnarray*}
\label{9}
&&R_{hijk, lm}-R_{hijk, ml}=A_{hm}R_{lijk}+A_{im}R_{hljk}+A_{jm}R_{hilk}+A_{km}R_{hijl}\nonumber\\&&
\,\,\,\,\,\,\,\,\,\,\,\,\,\,\,\,\,\,\,\,\,\,\,\,\,\,\,\,\,\,\,\,\,\,\,\,\,\,\,\,\,\,\,\,\,\,\,\,\,\,\,\,\,\,\,\,\,\,-A_{hl}R_{mijk}-A_{il}R_{hmjk}-A_{jl}R_{himk}-A_{kl}R_{hijm},
\end{eqnarray*}
where $A_{hm}=v_{h, m}-v_{h}v_{m}$. The foregoing equation together with equation (\ref{4}) and Ricci identity  entail that
\begin{eqnarray}
\label{10}
&&R_{pijk} R^{p}_{hlm}+R_{hpjk}R^{p}_{ilm}+R_{hipk}R^{p}_{jlm}+R_{hijp}R^{p}_{klm}\nonumber\\&&
\,\,\,\,\,\,\,\,\,\,\,\,\,\,\,\,\,\,\,=(\varphi g_{hm}+(\varphi -1)v_{h} v_{m})R_{lijk}+(\varphi g_{im}+(\varphi -1)v_{i}v_{m})R_{hljk}\nonumber\\&&
\,\,\,\,\,\,\,\,\,\,\,\,\,\,\,\,\,\,\,\,\,\,\,\,+(\varphi g_{jm}+(\varphi-1)v_{j}v_{m})R_{hilk}+(\varphi g_{km}+(\varphi-1)v_{k}v_{m})R_{hijl} \nonumber\\&&
\,\,\,\,\,\,\,\,\,\,\,\,\,\,\,\,\,\,\,\,\,\,\,\,-(\varphi g_{hl}+(\varphi-1)v_{h}v_{l})R_{mijk}-(\varphi g_{il}+(\varphi-1)v_{i}v_{l})R_{hmjk}\nonumber\\&&
\,\,\,\,\,\,\,\,\,\,\,\,\,\,\,\,\,\,\,\,\,\,\,\,-(\varphi g_{jl}+(\varphi-1)v_{j}v_{l})R_{himk}-(\varphi g_{kl}+(\varphi-1)v_{k}v_{l})R_{hijm}.
\end{eqnarray}
Multiplying equation (\ref{10}) with $v^{h}$ and using (\ref{6}), we conclude that
\begin{eqnarray*}
\label{}
L. H. S. &=&R^{p}_{ijk} R_{phlm} v^{h}+v^{h} R_{hpjk} R^{p}_{ilm} +v^{h}R_{hipk}R^{p}_{jlm}+v^{h}R_{hijp}R^{p}_{klm}\nonumber\\&=&
-R_{lijk} w_{m}-v_{p} R^{p}_{ijk} v_{l} w_{m} +w_{l} R_{mijk} +w_{l} v_{p} R^{p}_{ijk} v_{m}+w_{k} v_{j}v_{p}R^{p}_{ilm}\nonumber\\&&
-w_{j}v_{p}v_{k}R^{p}_{ilm}+w_{k}v_{i}v_{p}R^{p}_{jlm}-w_{p}R^{p}_{jlm}g_{ik}-w_{p}v_{i}v_{k}R^{p}_{jlm}\nonumber\\&&
+w_{p} R^{p}_{klm} g_{ij}+w_{p}v_{i}v_{j} R^{p}_{klm}-w_{j}v_{i} v_{p} R^{p}_{klm}.
\end{eqnarray*}
Again, using equations (\ref{6}) and $w_{k}=f v_{k}$ for some scalar $f$ in the foregoing equation we have
\begin{equation*}
\label{}
L. H. S. =-f v_{m} R_{lijk}+f v_{l} R_{mijk}-f^2 v_{i} v_{k}(v_{m}g_{jl }-v_{l}g_{jm})+f^2 (v_{m}g_{kl}-v_{l}g_{km})g_{ij}.
\end{equation*}
Following the above process, the right-hand side of equation (\ref{10}) assumes the form
\begin{eqnarray*}
\label{}
&&R. H. S.=v_{m} R_{lijk}-v_{l}R_{mijk}+f(\varphi-1) v_{i} v_{m}\{v_{k}g_{lj}-v_{j}g_{lk}\}\nonumber\\&&
\,\,\,\,\,\,\,\,\,\,\,\,\,\,\,\,\,\,\,\,\,\,\,\,-f \varphi g_{jm} v_{l} g_{ik}+f (\varphi -1 )v_{j}v_{m}\{v_{k}g_{il}-v_{l}g_{ik}\}\nonumber\\&&
\,\,\,\,\,\,\,\,\,\,\,\,\,\,\,\,\,\,\,\,\,\,\,\,+f (\varphi g_{km}+(\varphi-1)v_{k}v_{m})(v_{l}g_{ij}-v_{j}g_{il})+f \varphi v_{j}g _{il} g_{mk}\nonumber\\&&
\,\,\,\,\,\,\,\,\,\,\,\,\,\,\,\,\,\,\,\,\,\,\,\,-f(\varphi-1)v_{i} v_{l }(v_{k}g_{mj}-v_{j}g_{mk})+f \varphi g_{jl} v_{m} g_{ik}\nonumber\\&&
\,\,\,\,\,\,\,\,\,\,\,\,\,\,\,\,\,\,\,\,\,\,\,\,-f(\varphi-1)v_{j} v_{l}\{v_{k}g_{im}-v_{m}g_{ik}\}-f \varphi g_{kl} v_{m} g_{ij}\nonumber\\&&
\,\,\,\,\,\,\,\,\,\,\,\,\,\,\,\,\,\,\,\,\,\,\,\,-f (\varphi-1) v_{k} v_{l}\{v_{m}g_{ij}-v_{j}g_{im}\}.
\end{eqnarray*}
The above equations infer that
\begin{eqnarray}
\label{11}
&&-f v_{m} R_{lijk} +f v_{l} R_{mijk} -f^2 v_{i} v_{k} (v_{m}g_{il}-v_{l}g_{jm})+f^2 (v_{m} g_{kl}-v_{l}g_{km})g_{ij}\nonumber\\&&
 \,\,\,\,\,\,\,\,\,\,\,\,\,\,\,\,\,\,\,\,\,\,\,\,=v_{m}R_{lijk}-v_{l}R_{mijk}+f(\varphi-1)v_{i}v_{m}\{v_{k} g_{lj}-v_{j}g_{lk}\}\nonumber\\&&
\,\,\,\,\,\,\,\,\,\,\,\,\,\,\,\,\,\,\,\,\,\,\,\,\,\,\,\,\,-f \varphi g_{jm} v_{l} g_{ik}+f (\varphi -1 )v_{j}v_{m}\{v_{k}g_{il}-v_{l}g_{ik}\}\nonumber\\&&
\,\,\,\,\,\,\,\,\,\,\,\,\,\,\,\,\,\,\,\,\,\,\,\,\,\,\,\,\,+f (\varphi g_{km}+(\varphi-1)v_{k}v_{m})(v_{l}g_{ij}-v_{j}g_{il})+f \varphi v_{j}g _{il} g_{mk}\nonumber\\&&
\,\,\,\,\,\,\,\,\,\,\,\,\,\,\,\,\,\,\,\,\,\,\,\,\,\,\,\,\,-f(\varphi-1)v_{i} v_{l }(v_{k}g_{mj}-v_{j}g_{mk})+f \varphi g_{jl} v_{m} g_{ik}\nonumber\\&&
\,\,\,\,\,\,\,\,\,\,\,\,\,\,\,\,\,\,\,\,\,\,\,\,\,\,\,\,\,-f(\varphi-1)v_{j} v_{l}\{v_{k}g_{im}-v_{m}g_{ik}\}-f \varphi g_{kl} v_{m} g_{ij}\nonumber\\&&
\,\,\,\,\,\,\,\,\,\,\,\,\,\,\,\,\,\,\,\,\,\,\,\,\,\,\,\,\,-f (\varphi-1) v_{k} v_{l}\{v_{m}g_{ij}-v_{j}g_{im}\}.
\end{eqnarray}
Multiplying equation (\ref{11}) with $v^{l}$, and making use of equations (\ref{6}) and $w_{k}=f v_{k}$ we obtain
\begin{eqnarray*}
\label{}
&&-f^2 v_{m} (v_{k}g_{ij}-v_{j}g_{ik})-R_{mijk}-f^2 v_{i}v_{k}(v_{m}v_{j}+g_{jm})+f^{2}(v_{m}v_{k}+g_{mk})g_{ij} \nonumber\\&&
=f^2 v_{m} \{v_{k}g_{ij}-v_{j}g_{ik}\}+R_{mijk}+f(\varphi-1) v_{j} v_{m}\{v_{k}v_{i}+g_{ik}\}+f \varphi g_{jm}g_{ik}\nonumber\\&&
\,\,\,\,+f\{ \varphi g_{jm} +(\varphi-1) v_{j} v_{m}\}(v_{k}v_{i}+g_{ki})+f(\varphi-1)v_{i}(v_{k}g_{mj}-v_{j}g_{mk})\nonumber\\&&
\,\,\,\,+f \varphi v_{j} v_{i}g_{mk}+f \varphi v_{j} v_{m} g_{ik} +(\varphi -1) v_{j}\{v_{k}g_{im}-v_{m} g_{ik}\}\nonumber\\&&
\,\,\,\,-f \varphi v_{k} v_{m} g_{ij}+f (\varphi-1)v_{k}\{v_{m} g_{ij}-v_{j}g_{im}\}.
\end{eqnarray*}
Again multiplying the above equation with $g^{ij}$, we find
\begin{equation}
\label{2.1a}
R_{km}=\alpha g_{km}+\beta v_{k} v_{m},
\end{equation}
where $\beta=\frac{1}{2}[-f^2+f\{(2 \varphi-1)+n(\varphi-1)\}]$ and $\alpha=\frac{1}{2}[-n f^2+f(\varphi-1)]$. The last equation together with (\ref{pfs}) show that the pseudosymmetric $GRW$ spacetime is a perfect fluid spacetime. This finishes the proof of the Theorem.

\begin{remark}
\label{rm1}
In \cite{manticamolinaride}, Mantica, Molinari and first author proved that a perfect fluid spacetime in dimension $n \ge 4$, with differentiable equation of state $p=p(\sigma)$, $p+\sigma\ne 0$, and with null divergence of the Weyl conformal curvature tensor  $( C^{m}_{jkl, m}=0)$, is a $GRW$ spacetime. In Theorem \ref{thm1}, we gave an affirmative answer of the question "Under which condition a $GRW$ spacetime to be a perfect fluid spacetime?"
\end{remark}

In consequence of equations (\ref{2}), (\ref{3a}) and (\ref{2.1a}), we notice that
\begin{equation*}
\label{}
\left(\alpha-\frac{R}{2}-\kappa p \right)g_{ij}+(\beta-\kappa(p+\sigma)) v_{i}v_{j}=0,
\end{equation*}
which gives
$$ \kappa \sigma=(n-1)\alpha \,\,\,\,\, \text{and} \,\,\,\, (n-1)p \kappa=(n-1)\alpha -\frac{(n-2)(n \alpha-\beta)}{2}.$$
These reflect that the equation of state assumes the form
\begin{equation}
\label{2.1b}
\frac{p}{\sigma}=\frac{1}{n-1}-\frac{(n-2)(n \alpha-\beta)}{2\alpha (n-1)^2}.
\end{equation}
For $n=4$, the above equation with (\ref{2.1a}) take the form
\begin{equation*}
\label{}
\frac{p}{\sigma}=\frac{1}{3}+\frac{15f+2 \varphi-1}{\varphi-4f-1}.
\end{equation*}
If $2 \varphi=1-15f$, then the above equation becomes $3p=\sigma$, that is, the equation of state represents the radiation era in the evolution of the universe \cite{wein}. Now, we state our finding as:
\begin{corollary}
Let a pseudosymmetric $GRW$ spacetime satisfy the Einstein's field equations without cosmological constant, then the equation of state is given by (\ref{2.1b}). Also, if $n=4$ and $2\varphi =1-15f$, the matter of the spacetime represents the radiation era.
\end{corollary}

\section{Conformally flat $(PS)_n$ spacetimes}
This section deals with the study of pseudosymmetric conformally flat spacetimes. {\vspace{0.2cm}}\\
 {\it Proof of Theorem \ref{thm2}.} It is well-known that in a conformally flat $(PS)_{n}$ spacetime, the Ricci tensor $R_{ij}$ is of the form
\begin{equation*}
\label{3.1}
R_{ij}=\frac{R-t}{n-1}g_{ij}+\frac{nt-R}{(n-1)v_{p}v^{p}} v_{i} v_{j},
\end{equation*}
where $R$ denotes the scalar curvature and $t$ is a scalar \cite{tarafdar1}. The above equation can be rewritten as
\begin{equation}
\label{3.2}
R_{ij}=\alpha g_{ij}+ \beta \lambda_{i} \lambda_{j},
\end{equation}
where $\alpha=\frac{R-t}{n-1}$ and $\beta=\frac{nt-R}{(n-1)v_{p}v^{p}}$ are two scalars and $\lambda_{i}=\frac{v_{i}}{\sqrt{-v_{j} v^{j}}}$ is a unit timelike vector. Equation (\ref{3.2}) reflects that the conformally flat $(PS)_n$ spacetimes are the perfect fluid spacetimes.

According to our hypothesis, the conformal curvature tensor $C$ of our ambient space vanishes and therefore $div \, C=0$, where $div$ denotes the divergence. In \cite{manticamolinaride}, Mantica, Molinari and De proved that a perfect fluid spacetime with $div \, C=0$ is a $GRW$ spacetime. Thus, by considering equation (\ref{3.2}) and the above discussions, we obtain the required result. 


\section{Viscous fluid}
We consider the pseudosymmetric spacetimes with constant scalar curvature. From equation (\ref{1}) we get
\begin{equation*}
R_{ij, l}=2v_{l}R_{ij}+v_{i}R_{lj}+v_{j}R_{il}+v^{k}R_{lijk}+v^{k}R_{kijl}.
\end{equation*}
Considering an orthonormal frame field and contracting the above equation entails that $R_{, l}=2v_{l}R+4R_{lh}v^{h}$. Since the scalar curvature $R$ is constant,  therefore we get $R_{lh}v^{h}=-\frac{R}{2}v_{l}$, which implies that $-\frac{R}{2}$ is an eigenvalue of the Ricci tensor $R_{lh}$ corresponding to the eigenvector $v_{l}$.

In viscous fluid, the energy momentum tensor \cite{mno} is given by
\begin{equation}
\label{4.1}
T_{ij}=(p+\sigma)v_{i}v_{j}+p g_{ij}+D_{ij},
\end{equation}
where $D_{ij}$ is a $(0,2)$-type symmetric tensor, named as the anisotropic pressure of the fluid. Also $D_{ij}$ satisfies $D_{ij}v^{j}=0$ and trace of $D=0$. Suppose the spacetime under consideration obeys Einstein's field equation without cosmological constant, that is equation (\ref{3a}) is satisfied. Then we have
\begin{equation}
\label{4.2}
R_{ij}-\frac{R}{2}g_{ij}=\kappa [(p+\sigma)v_{i}v_{j}+p g_{ij}+D_{ij}].
\end{equation}
Multiplying equation (\ref{4.2}) with $g^{ij}$ we find
\begin{equation}
\label{4.3}
\frac{2-n}{2\kappa}R=(n-1)p-\sigma,
\end{equation}
since trace of $D=0.$ Now, multiplying equation (\ref{4.2}) with $v^{j}$ we infer
\begin{equation}
\label{4.4}
-R v_{i}=-\kappa \sigma v_{i} \implies \kappa \sigma =R,
\end{equation}
since $D_{ij}v^{j}=0$ and $R_{ij}v^{j}=-\frac{R}{2}v_{i}$. Equations (\ref{4.3}) and (\ref{4.4}) entail that
\begin{equation}
\label{4.5}
p=\frac{4-n}{2(n-1)}\sigma,
\end{equation}
which means that the fluid is isentropic \cite{manticasuhde}. Thus, we can write our result as:
\begin{theorem}
\label{thm4.1}
Let the pseudosymmetric spacetime with the constant scalar curvature satisfy the Einstein's field equations. If the spacetime is filled with viscous fluid, then the fluid is isentropic and the equation of state is given by (\ref{4.5}).
\end{theorem}

\begin{remark}
For $n=4$, equation (\ref{4.5}) reflects that $p=0$, that is the isotropic pressure of the viscous fluid is zero and the spacetime is filled with dust matter.
\end{remark}

In case of dust or pressure-less fluid, the energy momentum tensor assumes the form
\begin{equation}
\label{4.6}
T_{ij}=\sigma v_{i} v_{j}.
\end{equation}
In consequence of equation (\ref{3a}), equation (\ref{4.6}) takes the form
\begin{equation}
\label{4.7}
R_{ij}-\frac{R}{2}g_{ij}=\kappa \sigma v_{i} v_{j}.
\end{equation}
Let us consider an orthonormal frame field and contracting the above equation for $i$ and $j$, we get
\begin{equation}
\label{4.8}
\frac{2-n}{2}R=-\kappa \sigma.
\end{equation}
Again, multiplying equation (\ref{4.7}) with $v^{j}$ we lead to
\begin{equation}
\label{4.9}
-R v_{i}=-\kappa \sigma v_{i} \implies  \kappa \sigma=R,
\end{equation}
since $R_{ij} v^{j}=-\frac{R}{2}v_{i}$. From equations (\ref{4.8}) and (\ref{4.9}) we infer that $\sigma=0$ and hence equation (\ref{4.6}) reduces to $T_{ij}=0$, which implies that the spacetime is vacuum.
\begin{theorem}
\label{thm4.2}
Let a pseudosymmetric spacetime with the constant scalar curvature satisfy the Einstein's field equation without a cosmological constant. If the spacetime is filled with dust fluid, then it is vacuum.
\end{theorem}

\section{An Example of Pseudosymmetric spacetime}
In this section, we construct a non-trivial example to prove the existence of a pseudosymmetric spacetime whose associated vector is irrotational.

Let us consider a Lorentzian metric $g$ on $\mathbb{R}^4$ by
$$ds^2=g_{ij}dx^i dx^j=(dx^1)^{2}+(x^1)^{2}(dx^2)^{2}+(x^2)^2 (dx^3)^2-(dx^4)^2,$$
where $(x^1, x^2, x^3, x^4)$ are the standard coordinate of $\mathbb{R}^4$ and $i, j=1, 2, 3, 4$.
The only non-vanishing components of the Christoffel symbols and curvature tensor are
$$\Gamma^{1}_{22}=-x^1, \,\,\, \Gamma^{2}_{33}=-\frac{x^2}{(x^1)^2}, \,\,\, \Gamma^2_{12}=\frac{1}{x^1}, \,\,\, \Gamma^{3}_{23}=\frac{1}{x^2} \,\, \text{and}\,\, R_{1332}=-\frac{x^2}{x^1}.$$
From equation $R_{hijk, l}=2v_{l}R_{hijk}+v_{h}R_{lijk}+v_{i}R_{hljk}+v_{j}R_{hilk}+v_{k}R_{hijl}$ we have
\begin{equation}
\label{i}
R_{1332,1}=2v_{1}R_{1332}+v_{1}R_{1332}+v_{3}R_{1132}+v_{3}R_{1312}+v_{2}R_{1331},
\end{equation}
\begin{equation}
\label{ii}
R_{1332,2}=2v_{2}R_{1332}+v_{1}R_{2332}+v_{3}R_{1232}+v_{3}R_{1322}+v_{2}R_{1332},
\end{equation}
\begin{equation}
\label{iii}
R_{1332,3}=2v_{3}R_{1332}+v_{1}R_{3332}+v_{3}R_{1332}+v_{3}R_{1332}+v_{2}R_{1333},
\end{equation}
\begin{equation}
\label{iv}
R_{1332,4}=2v_{4}R_{1332}+v_{1}R_{4332}+v_{3}R_{1432}+v_{3}R_{1342}+v_{2}R_{1334}.
\end{equation}
We choose the covariant vector $v_i$ as follows:\\
\begin{equation*}
v_{i}=\left\{ \begin{array}{rcl}
-\frac{2}{3x^1}, & \mbox{for}
& i=1 \\ \\-\frac{1}{3x^2}, & \mbox{for} & i=2 \\\\
0, & \mbox{for} & i=3, 4.
\end{array}\right.
\end{equation*}
It can be easily shown that
\begin{eqnarray*}
\label{v}
&&v_{1, 4}=v_{4, 1}=0, \,\, v_{1, 2}=v_{2, 1}=\frac{1}{3x^1 x^2},\,\, v_{1, 3}=v_{3, 1}=0, \nonumber\\&&
\,\,\,\,\, v_{2, 4}=v_{4, 2}=0, \,\, v_{2, 3}=v_{3, 2}=0, \,\, v_{3, 4}=v_{4, 3}=0.
\end{eqnarray*}
From the above discussions, we can conclude that the equations (\ref{i})-(\ref{iv}) are true. Thus, the manifold under consideration is a pseudosymmetric spacetime. Moreover, equation (\ref{5}) shows that the covector ($1$-form) $v$ is irrotational (closed).

\end{document}